\documentclass[10pt]{article}
\usepackage{graphicx,xspace,colortbl}
\usepackage{makeidx}

\makeindex
\makeglossary

\usepackage{amsmath,amsthm,amsfonts}
\usepackage{fancybox}

\usepackage{hyperref}

\nonstopmode

\usepackage{tikz}
\usepackage[round]{natbib}
\usepackage{amsmath}
\usepackage{amsbsy}
\usepackage{amscd}
\usepackage{amsopn}
\usepackage{amstext}
\usepackage{amsxtra}
\usepackage{amssymb}
\usepackage{amsthm,amsfonts}
\usepackage{graphicx}
\usepackage{epstopdf}
\usepackage{bbm}
\usepackage{epsfig}
\usepackage{cancel}
\usepackage{multirow}
\usepackage[margin=0.5in,font=small]{caption}
\usepackage{subcaption}
\usepackage{centernot}
\usepackage{color}
\usepackage{sectsty}
\usepackage[title,titletoc]{appendix}
\usepackage{titling}

\sectionfont{\fontsize{10}{15}\selectfont}
\subsectionfont{\fontsize{10}{15}\selectfont\textit}
\subsubsectionfont{\fontsize{10}{15}\selectfont\it}

\def\ls{\left[}
\def\rs{\right]}

\def\r{\rho}
\def\dx{\mathrm{d}x}
\def\dy{\mathrm{d}y}

\def\CB{{\cal B}}

\def\CD{{\cal D}}
\def\CE{{\cal E}}
\def\CF{{\cal F}}
\def\CG{{\cal G}}

\def\CM{{\cal M}}
\def\CN{{\cal N}}

\def\CS{{\cal S}}

\def\BBE{\mathbb{E}}
\def\BBP{\mathbb{P}}
\def\BBN{\mathbb{N}}

\def\BBC{\mathbb{C}}
\def\dotstar{\cdot\! \ast}
\def\BBR{\mathbb{R}}

\def\l{\left}
\def\r{\right}

\def\WOR{{~\centernot{\circledR}~}}

\def\Nvp{{\CN_v^+}}
\def\Nvm{{\CN_v^-}}

\def\1{\mathbbm{1}}

\def\beq{\begin{eqnarray}}
\def\eeq{\end{eqnarray}}
\def\beqq{\begin{eqnarray*}}
\def\eeqq{\end{eqnarray*}}
\def\beeq{\begin{eqnarray*}}
\def\eeeq{\end{eqnarray*}}
\def\be{\begin{equation}}
\def\ee{\end{equation}}
\renewenvironment{abstract}
 {\small
  \begin{center}
  \bfseries \abstractname\vspace{-.5em}\vspace{0pt}
  \end{center}
  \list{}{
    \setlength{\leftmargin}{0.25in}%
    \setlength{\rightmargin}{\leftmargin}%
  }%
  \item\relax}
 {\endlist}

\theoremstyle{remark}

\newtheorem{definition}{Definition}
\newtheorem{remark}{Remark}

\makeindex             

\numberwithin{equation}{section}
\begin{document}

\title{\normalsize \bf DOUBLE CASCADE MODEL OF FINANCIAL CRISES}
\author{\footnotesize T. R. HURD \vspace{-4ex}\\
\footnotesize \it Mathematics \& Statistics, McMaster University, 1280 Main St. West \vspace{-5ex}\\
\footnotesize \it Hamilton, Ontario, L8S4L8, Canada \vspace{-5ex}\\
\footnotesize \it hurdt@mcmaster.ca\\
\and
\footnotesize DAVIDE CELLAI \vspace{-4ex}\\
\footnotesize \it MACSI, Department of Mathematics and Statistics, University of Limerick \vspace{-5ex}\\
\footnotesize \it Limerick, V94 T9PX, Ireland \vspace{-5ex}\\
\footnotesize \it davide.cellai@gmail.com\\
\and
\footnotesize SERGEY MELNIK \vspace{-4ex}\\
\footnotesize \it MACSI, Department of Mathematics and Statistics, University of Limerick \vspace{-5ex}\\
\footnotesize \it Limerick, V94 T9PX, Ireland \vspace{-5ex}\\
\footnotesize \it sergey.melnik@ul.ie\\
\and
\footnotesize QUENTIN H. SHAO\vspace{-4ex}\\
\footnotesize \it Mathematics \& Statistics, McMaster University, 1280 Main St. West \vspace{-5ex}\\
\footnotesize \it Hamilton, Ontario, L8S4L8, Canada \vspace{-5ex}\\
\footnotesize \it shaoq@math.mcmaster.ca}
\predate{\footnotesize\centering Received (day,month,year)\\}
\postdate{\footnotesize\centering Revised (day,month,year)\\}
\date{}
\maketitle

\begin{abstract}
\footnotesize
\setlength{\baselineskip}{10pt}
The scope of financial systemic risk research encompasses a wide range of interbank channels and effects, including asset correlation shocks, default contagion, illiquidity contagion, and asset fire sales. This paper introduces a financial network model that combines the default and liquidity stress mechanisms into a ``double cascade mapping''. The progress and eventual result of the crisis is obtained by iterating this mapping to its fixed point. Unlike simpler models, this model can therefore quantify how illiquidity or default of one bank influences the overall level of liquidity stress and default in the system. Large-network asymptotic cascade mapping formulas are derived that can be used for efficient network computations of the double cascade. Numerical experiments then demonstrate that these asymptotic formulas agree qualitatively with Monte Carlo results for large finite networks, and quantitatively except when the initial system is placed in an exceptional ``knife-edge'' configuration. The experiments clearly support the main conclusion that when banks respond to liquidity stress by hoarding liquidity, then in the absence of asset fire sales, the level of defaults in a financial network is negatively related to the strength of bank liquidity hoarding and the eventual level of stress in the network.\\
{\bf Keywords:} Systemic risk, banking network, contagion, random graph, default, funding liquidity, liquidity hoarding.
\end{abstract}
\renewcommand{\thefootnote}{\alph{footnote}}

\section{Introduction}

Since the banking crisis of 2007-2008, the types of shocks transmitted through interbank networks during a crisis  now thought to be important include not only shocks arising from defaulting banks, but a number of additional phenomena, most notably asset shocks originating with forced sales by banks of illiquid assets and funding liquidity shocks as illiquid banks recall loans from other banks. 
A comprehensive treatment of these contagion effects and how they can be computed in stylized financial network specifications can be found in \cite{Hurd2016}.

A well-developed strand of literature on bank default cascade models, starting with \cite{EiseNoe01}  and reviewed in \cite{Upper11}, is based on a network of banks wherein insolvency of a given bank, defined as a bank whose net worth becomes non-positive, will generate shocks to the asset side of the balance sheet of each of its creditor banks. Under some circumstances, such ``downstream'' shocks can cause further insolvencies that may build up to create a global insolvency cascade. One such contribution is by \cite{NieYanYorAle07} that uses Monte Carlo simulation to determine how the key network parameters for their model of defaults in a stylized random network of 25 banks can influence the total number of defaults in a nonlinear, indeed sometimes nonmonotonic, fashion.  The paper of \cite{GaiKapa10a} and its extension by \cite{HurdGlee11}, adapt the \cite{Watts02} network model of information cascades to the context of financial systems, deriving both analytical and Monte Carlo results for the dependence of the default cascade on different structural parameters. \cite{MayArin10} present approximate analytical formulas for the Nier et al model and Gai-Kapadia model that can explain some of the main properties of the simulation results found in those papers. \cite{AminContMinc12} and \cite{AminContMinc16} develop a simple but general analytical criterion for resilience to default contagion, based on an asymptotic analysis of default cascades in heterogeneous networks. 

More recently, after remarking on the observed ``freezing'' of interbank lending around the time of the Lehman collapse, papers by \cite{GaiKapa10b}, \cite{GaiHalKap11} and \cite{SHLee13} adopt variations of an idea that funding illiquidity\footnote{Funding illiquidity, being the insufficiency of liquid assets to cover a run on liabilities,  is distinct from market illiquidity, where assets become difficult to sell due to an oversupply in the market. See \cite{Tirole10} and \cite{BrunPede09}  for a detailed analysis of these concepts.}  of a bank can also be transmitted contagiously through interbank exposures. They argue that a bank whose liquid assets are insufficient to cover demands on its liabilities will be illiquid or ``stressed'', and must reduce its interbank lending, thereby creating shocks to the liability side of its debtor banks' balance sheets. Such ``upstream'' shocks can cause illiquidity stress in other banks that in some circumstances may build up to create a global illiquidity cascade.  

A third channel, sometimes called market illiquidity or the asset fire sale effect,  is identified by \cite{CifuFerrShin05} who consider a network of banks that hold a common asset. As stress in the system builds up, some banks are forced to sell large positions in the illiquid asset, leading to a downward price spiral coupled to worsening bank balance sheets across the network. This mechanism has been extended to the sharing of multiple assets in a paper by \cite{CacShrMooFar14}. 

These papers  simplify by omitting economic details that obscure their focus on a single channel of systemic risk. They share the view that the end result of a crisis is a new equilibrium that is reached through a sequence of mechanistic steps where banks update and modify their balance sheets in response to price changes and shocks transmitted from other banks.  Usually, properties of the system at step $n$ of the cascade can be computed in terms of the  properties at step $n-1$, yielding a set of equations that can be called a {\it cascade mapping}. In some  models, an analytic cascade mapping can be found that yields a detailed picture of systemic risk without the need for Monte Carlo simulation based computations. Even in such cases however,  large scale benchmarking of the analytic approximation method against Monte Carlo simulations is undertaken to validate its use. 

\begin{figure}[!htbp]
\centering\includegraphics[%
  scale=.6]{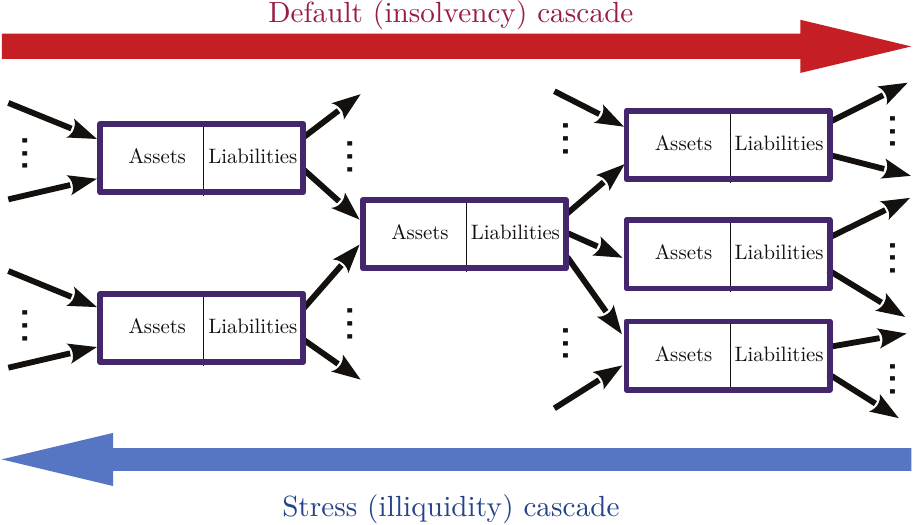}
   \caption{A schematic diagram of part of a financial network, showing banks and their balance sheets connected by directed edges representing interbank exposures. Default shocks are transmitted in the forward ``downstream'' direction, while liquidity or stress shocks are transmitted in the upstream direction. }
   \label{networksketch}
\end{figure}

An important question now arises whether one can successfully integrate two or more such cascade mechanisms, incorporate ``spillover'' effects in a realistic, analytically tractable way, and discern new features of systemic risk. The purpose of the present paper is to do exactly this, by introducing a network model  of a double illiquidity-insolvency cascade, and deriving an analytic cascade mapping that describes it. This double cascade model, shown schematically in Figure \ref{networksketch}, can resolve issues not addressable within a single cascade model, for example, to quantify the effect of a bank's behavioural response to liquidity stress on the level of eventual defaults in the entire system. In particular, one can show that a  bank that reacts to stress at a time of crisis by shrinking its own interbank lending, thereby inflicting liquidity shocks to its debtor banks, will also protect itself from default due to interbank default shocks.

In the literature on how contagion channels ``spill over'' into each other, \cite{DiamRaja05} explain the liquidity, solvency and behaviour of banks through the role they play in intermediating between lenders and borrowers. As discussed in \cite{Tirole10}, the holding of very illiquid assets can cause solvent banks (those with positive equity) to default due to their inability to raise enough liquidity to meet short-term needs.   \cite{BaGaGaGrSt12} model the robustness dynamics of a network of banks linked by their interbank exposures, highlighting the feedback effects of a ``financial accelerator'' mechanism on the level of systemic risk. \cite{RoBePiCaBa13} further develop  this idea, investigating the systemic impact of network topologies and different levels of market illiquidity and applying this methodology to the Italian interbank money market from 1999 to 2011.

It should be made  clear that in order to focus on pure contagion effects, the double cascade model and the above models simplify by explicitly ruling out certain other systemic mechanisms that have been explored in the economics and finance literature. In the present paper, the most important economic effect that is ruled out is the asset ``fire sale'',  which means that for the duration of the crisis, changing values of fixed assets are ignored.  Some of the above references discuss the impact of financial cascades on the non-financial economy and the consequent feedback into the financial markets. They find that fire sales of assets amplify any cascade that takes hold in the network. It will be an important future challenge to extend our double cascade framework to include the fire sale effect and determine whether our main conclusions are robust to such an extension.


Section 2 of the paper sets out the network framework and assumptions underlying the balance sheet structure of banks, the timing of the crisis and bank behaviour. These  assumptions lead to stock flow consistent rules for the transmission of shocks through the double stress/default cascade, including the conditions for banks to become stressed or defaulted. Section 3 develops our main technical contribution, which yields an explicit analytic cascade mapping for default and stress probabilities valid in a large heterogeneous network of banks with random balance sheets and interbank exposures and connectivity given by a random ``skeleton''. Section 4 provides a parallel development of default and stress probabilities for cascades on networks where it is assumed that the skeleton of interbank connections is known explicitly, but balance sheets and exposure sizes are random. Several representative financial experiments are reported in Section 5. First we summarize experiments that validate the accuracy of the cascade mapping by direct comparison of large network analytics to Monte Carlo simulation results. Secondly, we investigate the relationship between the level of stress and default, verifying our assertion that average default probability decreases as average stress probability  increases. A final experiment  demonstrates that our analysis is still useful when pushed to a highly heterogeneous specification of the model that is consistent with known heuristics of financial networks, such as fat-tailed degree and exposure distributions, and the 2011 stress testing data on 90 large banks in the EU system.  We observed that the stylized EU network is very resilient, and only extremely large shocks to the average default buffer size will trigger a cascade of defaulted and stressed banks. 

One conclusion of this paper is that our analytic asymptotic results on default and stress probabilities are broadly consistent with results from Monte Carlo simulations on random networks.  A second conclusion is that under the assumption of no asset fire sales, stress and default are inversely related: as banks respond to stress more vigorously, creating more network stress, they protect the network from default. It is left for future work to determine the extent to which this conclusion is valid when a fire sale mechanism provides an additional channel causing stressed but solvent banks to default.

\section{Cascade Mechanisms}

The assumptions we now make lead to a stylized network model admitting double cascades of illiquidity and insolvency that can be computed efficiently under a wide range of initial conditions. The picture to adopt is that of a system hit ``out of the blue'' by a large shock that triggers a systemic crisis that proceeds stepwise in time.  Many realistic effects that would significantly complicate our model are left out. For example, banks respond mechanistically according to simple stylized rules rather than dynamically optimizing their behaviour. We rule out other channels of systemic risk, in particular asset fire sales. We take a static view, and ignore external cash flows, interest payments and price changes for the duration of the cascade. We make an assumption of zero-recovery on interbank assets on the default of a counterparty during the cascade. Our assumptions imply that insolvency (that is, when equity is nonpositive) is equivalent to default (that is, the bank can no longer pay some liability). This means that even when stressed, banks always pay recalled interbank liabilities in full, until the moment they become insolvent.
 
\begin{figure}
\centering\includegraphics[width=0.55\columnwidth]{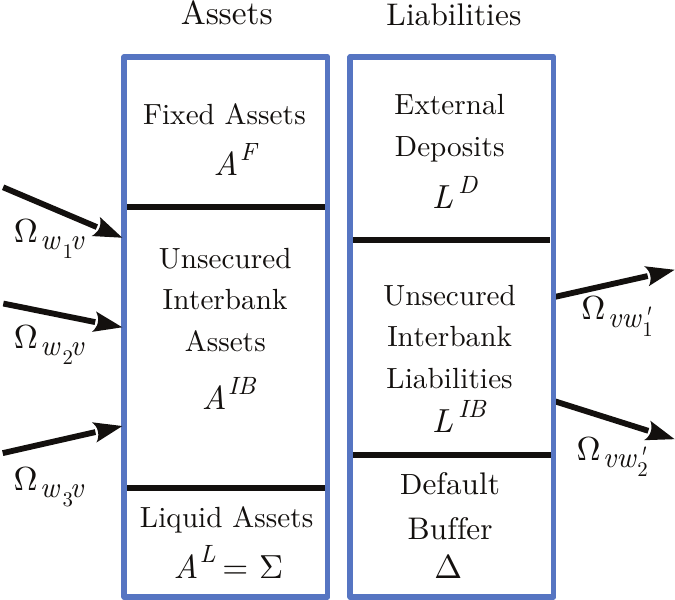}
   \caption{The stylized balance sheet of a bank $v$ with in-degree $j_v=3$ and out-degree $k_v=2$. Banks $w_1,w_2,w_3$ are debtors of $v$ while $w_1',w_2'$ are its creditors.
   The total exposure of $v$ to $w_1$ is denoted with $\Omega_{w_1v}$, and so on.
   The default buffer $\Delta$ of bank $v$ is the difference between assets and liabilities, and the ``stress buffer'' $\Sigma$ is the preferred asset class from which it pays creditors' demands.
   }
\label{balancesheet}
\end{figure}

The network consists of a collection of $N$ banks, labelled by numbers from the set $\CN=\{1,2,\dots, N\}$, each structured with a balance sheet as shown in Figure \ref{balancesheet}.

The subset of debtor banks of a bank $v\in\CN$ is called the in-neighbourhood of $v$, denoted by $\CN_v^-$, and the {\it in-degree} is the number of debtors $j_v=|\CN^-_v|$ .
Similarly the bank's creditor banks form a subset called the out-neighbourhood  $\CN^+_v$ of $v$ whose size $k_v=|\CN^+_v|$ is called the {\it out-degree}. The web of interbank counterparties is called the ``skeleton'', and is identifiable as a directed graph, i.e.  a collection of directed arrows, called edges, between pairs of nodes. Each debtor-creditor pair $v,w$ with $w\in\CN^+_v$ is denoted by an edge $\ell=(vw)$ pointing from $v$ to $w$.\footnote{The convention that arrows point from debtors to creditors means that default shocks propagate in the downstream direction.  Confusingly, much of the systemic risk literature uses the reverse convention that arrows point from creditors to debtors.} The {\it type} $(j,k)$ of a bank $v$ is its in-degree $j=j_v$ and out-degree $k=k_v$, and we write $v\in\CN_{jk}$. The {\it type} $(k,j)=(k_\ell, j_\ell)$ of an edge $\ell=(vw)$ is the out-degree $k_\ell=k_v$ of the debtor bank $v$ and the in-degree $j_\ell=j_w$ of the creditor bank $w$, and we write $\ell\in\CE_{kj}$.

The bank equity or net worth $\Delta_v=\max(A^{F}_v+A^{IB}_v+A^L_v-L^D_v-L^{IB}_v,0)$ of bank $v$ incorporates {\it limited liability}, which assumes that the bank must default and be forced to liquidate at the first moment it becomes insolvent, that is, its liabilities exceed its assets. Thus  $\Delta_v$ can be interpreted as the bank's {\it default buffer} that must always be kept positive to avoid default. 

Banks are also concerned about the possibility of runs on their liabilities, and try to keep a positive buffer of liquid assets $A^L$ (such as cash and government bonds), henceforth called the {\it stress buffer} $\Sigma$, which are the preferred asset class from which to pay creditors' demands.  A bank $v$ with $\Sigma_v$  fully depleted to zero will be called {\it stressed}. This does not mean the bank is {\it illiquid} in the sense of being unable to meet the creditors' demands. Rather it means that the bank is experiencing a significant degree of stress in meeting these demands and must turn to other assets, first interbank assets, then fixed assets, to realize the needed cash. The no fire sale assumption means that as long as they are solvent, stressed banks always meet their payment obligations.  

The interbank liabilities $L^{IB}$ and assets $A^{IB}$ decompose into bilateral interbank exposures. For any bank $v$ and one of its creditors $w\in\CN_v^+$, we denote  by $\Omega_{vw}$ the total exposure of $w$ to $v$. Then we have the constraints
\be A^{IB}_v=\sum_{w\in\CN_v^-} \Omega_{wv};\quad L^{IB}_v=\sum_{w\in\CN_v^+} \Omega_{vw} \ .\ee

Prior to the onset of the crisis, all banks are assumed to be in the normal state. Then, on day $n=0$, a  collection of banks, possibly all banks,  are assumed to experience initial shocks. Two kinds of initial shocks are possible. First, an asset shock causes a drop in the mark-to-market value of the fixed asset portfolio,  and reduces the default buffer. If the downward asset shock leads to $\Delta\le 0$, the bank must default.  The second kind of shock is a demand shock by the external depositors that reduces the stress buffer. If the demand shock leads to $\Sigma\le 0$, the bank becomes stressed. 

\bigskip
\noindent{\bf  Timing and Bank Behaviour Assumptions:\footnote{Treating cascade steps as daily is simply an aid to understanding.}\ }\begin{enumerate}
  \item  Prior to the crisis, all banks are in the normal state, neither stressed nor insolvent. The crisis commences on day $0$ after initial shocks trigger  the default or stress of one or more banks;
  \item Balance sheets are recomputed daily on a mark-to-market basis and banks respond daily on the basis of their newly computed balance sheets. All external cash flows, interest payments, and asset and liability price changes are ignored throughout the crisis. 
  \item An insolvent bank, characterized by $\Delta=0$, is forced into liquidation by the regulator. At this moment, each of its creditor banks are obliged to write down its defaulted exposures to zero thereby experiencing a {\it solvency shock} that  reduces its  default buffer $\Delta$. 
    \item A stressed bank, defined to be a non-defaulted bank with $\Sigma=0$, reacts one time only, at the moment it becomes stressed, by reducing its interbank assets $A^{IB}$ to $(1-\lambda)A^{IB}$ where $\lambda$ taken to be a fixed constant across all banks during the crisis. It does so by recalling a fraction $\lambda$ of its interbank loans, thereby transmitting a  {\it stress shock} to the liabilities each of its debtor banks.  Since nondefaulted banks are able to pay all of their liabilities, such recalled loans are repaid in full. 
    \item A newly defaulted bank also triggers maximal stress shocks (i.e. with $\lambda=1$) to each of its debtor banks as its bankruptcy trustees recall all its interbank loans, reducing $A^{IB}$ to $0$;
  \item Stress shocks reduce any bank's stress buffer $\Sigma$ and stressed banks remain stressed until the end of the cascade or until they default. 
  \end{enumerate}

\begin{remark}
The stress  parameter $\lambda$ was introduced by \cite{GaiHalKap11} to simplify banks' response to liquidity stress and to capture the effect of liquidity hoarding. We suppose that banks react sufficiently strongly to stress by choosing $\lambda\in [ 0.4, 1]$, to capture the picture that they hoard liquidity to preempt the need for further response later in the crisis. Unlike default, a bank is free to set its response to liquidity: both $\lambda$ and the size of the stress buffer $\Sigma$ are its own policy decisions. In a more realistic and complex model, $\lambda$ would be endogenously determined for each bank, reflecting the cumulative demands on its liabilities.  \end{remark}

The dynamics of the cascade that follow from these assumptions is stock flow consistent. This means that as long as both counterparties are solvent, every transaction involves four equal and opposite balance sheet adjustments: to both the assets and liabilities of both counterparties. The dynamics also has the property that it is completely determined by a reduced set of balance sheet data that consists of the collection of buffers $\Delta, \Sigma$ and interbank exposures $\Omega$. 

We shall now apply the cascade rules to a financial network found on day $0$ of the crisis  in an initial state  described by the following  elements: the interbank links which form a directed graph $\CE$ on the set of banks $v\in \CN$, called the {\it skeleton}; the buffers $\Delta_v, \Sigma_v$ for nodes $v\in\CN$; and the exposures  $\Omega_{vw}$ for edges $\ell=(vw)\in\CE$. After $n$ cascade steps, we identify $\CD_n$, the set of  defaulted banks,  $\CS_n$, the set of undefaulted banks that are under stress after $n$ steps, and $\CD^c_n\cap\CS^c_n$ which contains the remaining undefaulted, unstressed banks.\footnote{For any set $\CB$, $\CB^c$ denotes its complement. For an {\it event} defined by some condition $P$, for example $v\in\CD_n$,  the indicator function for that event is written $\1(P)$.} In our model, banks do not recover from either default or stress during the crisis, so the sequences $\{\CD_n\}_{n\in\BBN}$ and $\{
\CD_n\cup\CS_n\}_{n\in\BBN}$ are non-decreasing. 

To say that a bank $v$ is defaulted at step $n$ means that default shocks to step $n-1$ exceed its default buffer:
\be
\displaystyle
\CD_n :=
		\begin{cases}
		\{\Delta_v = 0\} & \text{ for } n=0\ \\\\
		\Bigl\{v \  \mid  \ \sum_{w\in \Nvm} \Omega_{wv}\xi^{(n-1)}_{wv} \ge \Delta_v\Bigr\} & \text{ for } n \ge 1\ ,
		\end{cases}
		\label{Dndef}
\ee
where the variables $\xi$ indicate the fractional sizes of the various default shocks impacting $v$.  Similarly, for any upstream neighbour $w\in\CN^-_v$, we say that $v$ is defaulted at step $n$ {\it without regarding $w$} and write $v\in \CD_n \WOR w$ under the conditions
\be\label{default_WOR}
\displaystyle
 \CD_n \WOR w :=
		\begin{cases}
		\{\Delta_v = 0\}\cap\CN^+_w & \text{ for } n=0\  \\\\
		\Bigl\{v\in \CN^+_w\  \mid  \ \sum_{v'\in \Nvm\setminus w} \Omega_{v'v}\xi^{(n-1)}_{v'v} \ge \Delta_v\Bigr\} & \text{ for } n \ge 1\ .
		\end{cases}
\ee
Finally, to say $v$ is stressed at step $n$ means both that it is not yet defaulted and the stress shocks to step $n-1$ exceed the stress buffer, i.e. $\CS_n=\CD^c_n\cap\hat\CS_n$ where
\be
\label{stress}
\hat\CS_n :=
	\begin{cases}
	\{\Sigma_v = 0\} & \text{ for } n = 0\ \\\\
        \Bigl\{v\  \mid  \ \sum_{w\in \Nvp}\Omega_{vw}\zeta^{(n-1)}_{vw}\ge \Sigma_v \Bigr\} & \text{ for } n\ge 1\ ,
         \end{cases}
\ee
where the  variables $\zeta$ indicate the fractional sizes of the stress shocks impacting $v$. 

Accounting for the fact that at the moment when $v$ becomes stressed it reduces its interbank exposures, one can see that for $n\ge 1$ the fractions $\xi^{(n)}$ are given recursively by
\be
\label{xi_def}
\displaystyle
\xi_{wv}^{(n)} :=\begin{cases}
		\xi_{wv}^{(n-1)} & \text{ when } w\in \CD_{n-1}\cup\CD^c_{n}\ \\
1& \text{ when }  w\in\CD_{n}\setminus\CD_{n-1}\text{ and } v\in\hat\CS^c_{n-1}\  \\
1-\lambda& \text{ when }  w\in\CD_{n}\setminus\CD_{n-1}\text{ and } v\in\hat\CS_{n-1}\ , 
		\end{cases}
\ee
with the initial values  $\xi_{wv}^{(0)} = \1(w\in\CD_0)$. Similarly, accounting for the assumption that defaulted creditors transmit a maximal stress impact, whereas stressed creditors only transmit a stress shock of a fraction $\lambda$ of the interbank exposure, one has for $n\ge 0$
\be
\label{zetanew}
\displaystyle
\zeta_{vw}^{(n)} :=\begin{cases}
		0 & \text{ when } w\in \hat\CS^c_{n}\cap(\CD^c_{n} \WOR v)\ \\

\lambda& \text{ when }  w\in\hat\CS_{n}\cap(\CD_{n}^c \WOR v)\  \\
1& \text{ when }  w\in\CD_{n} \WOR v\ .
		\end{cases}
\ee

The reader will be curious to see the $\WOR$ condition in \eqref{zetanew}. The rationale is that the $\WOR$ condition explicitly  eliminates an apparent feedback where the stress shock to $v$ from $w$ at step $n$ seems to depend on the default state of $v$ at step $n-2$. This feedback is apparent, not real, because any stress shock to a defaulted bank is inconsequential.

The above prescription completely characterizes the cascade mapping of the system to itself. This mapping arrives at the fixed point that represents the end of the crisis in at most $2N$ steps when acting on a network of $N$ banks.
The next two sections are devoted to the derivation of two probabilistic versions of this double cascade mapping: first for the case of networks with random skeletons,  and second for the case of  networks with random balance sheets on a fixed skeleton.

\section{Networks with Random Skeletons}
Our model of a random financial network has three layers of structure: the skeleton (a random directed graph $(\CN,\CE)$);  the buffer random variables $\Delta_v,\Sigma_v$, $v\in\CN$ and the random exposures $\Omega_\ell,\ell\in \CE$.  We give a complete specification of the distributional properties of these random variables.

The skeleton $(\CN,\CE)$ is a random directed assortative configuration graph (ACG) of the type studied by \cite{Hurd16b},   generalizing the well-known undirected configuration random graph model introduced in \cite{BendCanf78} and \cite{Bollobas80}. 
The model is parametrized by $N$ and the node and edge type distributions 
\be P_{jk}= \BBP[v\in\CN_{jk}],\quad Q_{kj}=\BBP[\ell\in\CE_{kj}], \quad j,k\le K
\ee
where for simplicity in the following we assume in and out degrees $j,k$ are bounded by a constant $K$. $P$ and $Q$ can be considered as bivariate distributions which have marginals:
\be P^{+}_k:=\sum_j P_{jk}, \quad P^{-}_j:=\sum_kP_{jk},\quad Q^{+}_k:=\sum_j Q_{kj}, \quad Q^{-}_j:=\sum_kQ_{kj}\ .
\ee
{\it Assortativity} is defined to be the Pearson correlation of $Q$, considered as a bivariate probability distribution. Several studies of real financial networks, notably \cite{BechAtal10}, highlight the fact and relevance of their observed negative assortativity. 

\begin{definition}[The ACG Construction]\label{configdef} The node and edge type probability laws $P, Q$ are {\rm consistent} if:
\beq
z&:=&\sum_jjP^{-}_j=\sum_k kP^{+}_k\ ,\nonumber\\    Q^{+}_k&=&kP^+_k/z,\quad Q^{-}_j=jP^-_j/z, \mbox{ for all $j,k\le K$.}\eeq
Given consistent data $N,P,Q$,  a random graph $(\CN,\CE)$ with $N$ nodes is sampled as follows: \begin{enumerate}
  \item Draw a sequence of  $N$ node-type pairs $X= ((j_1,k_1),\dots, (j_N,k_N))$ independently from $P$, and accept the draw if and only if it is feasible, i.e. $\sum_{v\in \CN}\ j_v =\sum_{v\in \CN}\ k_v$, and this defines the number of edges $E$ that will result. Label the $v$th node with $k_v$ {\it out-stubs}\index{out-stub} (each out-stub is a half-edge with an out-arrow, labelled by its degree $k_v$) and $j_v$ {\it in-stubs}\index{in-stub}, labelled by their degree $j_v$.  Define the partial sums 
  \be  	u^-_j=\sum_v\mathbbm{1}(j_v=j),\quad
  	u^+_k=\sum_v\mathbbm{1}(k_v=k), \quad
 	u_{jk}=\sum_v\mathbbm{1}(j_v=j,k_v=k), 
  \ee
  the number $ e_k^+=k u_k^+$ of {\it k-stubs} (out-stubs of degree $k$)  and the number of {\it j-stubs} (in-stubs of degree $j$), $ e_j^-=j u_j^-$.
   \item Conditioned on $X$, the result of Step 1, choose an arbitrary ordering $\ell^-$ and $\ell^+$ of the $E$ in-stubs and $E$ out-stubs. The matching sequence, or ``wiring'', $W$ of edges  is selected by choosing a pair of permutations $\sigma,\tilde\sigma\in S(E)$ of the set $\CE$. This determines the edge sequence $\ell=(\ell^-=\sigma(\ell), \ell^+=\tilde\sigma(\ell))$   labelled by  $\ell\in\CE$, to which is attached a probability weighting factor
   \be\label{AWweighting}\prod_{\ell\in\CE} Q_{k_{\sigma(\ell)}j_{\tilde\sigma(\ell)}}\ .
   \ee
\end{enumerate}
  \end{definition}

The ACG simulation algorithm defines the required class of random graphs, but is infeasible because the acceptance condition in Step 1 is achieved only rarely, and drawing random permutations as in Step 2 is impractical.  The paper \cite{Hurd16b} proposes  an approximate simulation algorithm that appears to work well in practice. 

The non-negative default buffer random variables $\Delta_v, v\in\CN$ have point masses at $x=0$ that represent the bank initial default probability $p^0_v$. We assume that the distribution functions of $\Delta_v$ depend only on the type $(j,k)$, and have the following form:
\be\label{DeltaCDF} D_{jk}(x)=\BBP[\Delta_v\le x \mid v\in\CN_{jk}] \ ; \frac{d}{dx} D_{jk}(x):= p^0_{jk}\delta_0(x) + d_{jk}(x)\ .
\ee
where $d_{jk}(x)\ge 0$ is a specified function with $\int^\infty_0 d_{jk}(x)dx=1-p^0_{jk}$. 
Similarly, the stress buffer $\Sigma_v$ has a point mass at $x=0$ that represents this bank's initial stress probability $q^0_v$ and a distribution function that depends only on its node type $(j,k)$.  Thus  the stress buffer  distribution functions of nodes $v\in\CN_{jk}$ have the following form:
\be
\label{SigmaCDF} S_{jk}(x)=\BBP[\Sigma_v\le x,v\notin\CD_0 \mid v\in\CN_{jk}] \ ; \frac{d}{dx} S_{jk}(x):= q^0_{jk}\delta_0(x) + s_{jk}(x)\ .
\ee
where $s_{jk}(x)\ge 0$ is a specified function with $\int^\infty_0 s_{jk}(x)dx=1-p^0_{jk}-q^0_{jk}$. The exposure random variables $\Omega_\ell, \ell\in\CE$ are positive (i.e. there is zero probability to have a zero weight) and have distributions that depend only on the edge type $(k,j)$. These can be specified by the distribution functions 
\be\label{OmegaCDF}
W_{kj}(x)=\BBP[\Omega_\ell\le x \mid \ell\in\CE _{kj}];\ \frac{d}{dx} W_{kj}(x)=w_{kj}(x)\ .
\ee
Finally, conditional on the random skeleton  $(\CN,\CE)$,  the collection of random variables $\{\Delta_v,\Sigma_v, \Omega_\ell\}$ is assumed to be mutually independent. 

It was proven in \cite{Hurd2016} that the probability measure on random networks just defined has a property called locally tree-like independence (LTI)  extending the well-known locally tree-like property of configuration graphs that cycles of any fixed finite length occur with an asymptotically zero density as the number of nodes $N$ goes to infinity for fixed $P,Q$. The probabilistic analysis to follow rests on this extended type of independence that holds as $N\to\infty$: \\

\noindent {\bf The locally tree-like independence (LTI) property:\ } 
Consider the double cascade model defined by the collection of random variables $(\CN,\CE, \Delta, \Sigma, \Omega)$.
Let $\CN_1,\CN_2\subset\CN$ be any two subsets that share exactly one node $\CN_1\cap\CN_2=\{v\}$ and let $X_1,X_2$ be any pair of random variables where for each $i=1,2$, $X_i$ is determined by the information on $\CN_i$. Then $X_1$ and $X_2$ are conditionally independent, conditioned on the information $\Delta_v,\Sigma_v,j_v, k_v$ located at the node $v$.\medskip

The cascade mapping we now propose is based on the observation that a probability such as $\BBP[v\in\CD^n]$ that depends on the $n$th order neighbourhood of the node $v$ can be approximated iteratively, in terms of the probabilities of the states of the first order neighbouring nodes and edges of $v$ at step $n-1$. The accuracy of such a scheme depends on the degree of dependence between the states of these first order neighbours of $v$. In ideal situations, there is no dependence at all, and the cascade mapping is exact. In our present model, there are two sources of residual dependence amongst these neighbours, and our approximation amounts to neglecting this residual dependence. 

Let us define
\be
\begin{split}
	p_{jk}^{(n) } &= \BBP\l[v \in \CD_n \mid  v \in\CN_{jk}\r]\ ,\\
	q_{jk}^{(n)} &= \BBP\l[v \in \CS_n \mid  v \in\CN_{jk}\r]\ ,\\
	\widetilde p_{jk}^{(n) } &= \BBP\l[v \in \CD_n \WOR w \mid  v \in\CN_{jk}\cap\CN_w^+\r]\ ,\\
\label{hatqdef}	\widehat q_{jk}^{(n)} &=\BBP\l[v \in \hat\CS_n \mid  v \in\CN_{jk}\r]\  .
\end{split}
\ee
where $p_{jk}^{(n)}$ is the probability that a $(j,k)$-node has defaulted by time $n$, $q_{jk}^{(n)}$ is the probability that a $(j,k)$-node is stressed at time $n$, $\widetilde{p}_{jk}^{(n)}$ is the probability that a $(j,k)$-node has defaulted by time $n$, without regarding of one of its in-neighbours, and $\widehat q_{jk}^{(n)}$ is the probability that a $(j,k)$-node satisfies the stress condition \eqref{stress} at time $n$.
Our aim is to compute $p_{jk}^{(n)},\widehat q_{jk}^{(n)},\widetilde p_{jk}^{(n)} $, plus an auxiliary quantity 
\be t^{(n)}_{kj}=\BBP\l[\xi^{(n)}_{wv}=1 \mid  (w,v) \in\CE_{kj}\r],
\label{tdef}
\ee
recursively over $n$. It is also convenient to define 
\beq 
p^{(n)}_{k}&=&\BBP[v\in\CD_n \mid k_v=k]=\sum_{j}p^{(n)}_{jk}\ P_{j \mid k}\ ,\nonumber \\
\label{pphatq}\widetilde p^{(n)}_{j}&=&\BBP[v\in\CD_{n}\WOR w \mid j_v=j, v\in\CN^+_w]=\sum_{k}\widetilde p^{(n)}_{jk}\ P_{k \mid j}\ ,\\
 \widehat q^{(n)}_{j}&=&\BBP[v\in\hat \CS_{n} \mid j_v=j]=\sum_{k}\widehat q^{(n)}_{jk}\ P_{k \mid j}\ ,\nonumber
\eeq
where $P_{j \mid k}:=\frac{P_{jk}}{P^+_k}, P_{k \mid j}:=\frac{P_{jk}}{P^-_j}$.

Our computations will rely on two facts. The first is that if $X,Y$ are two independent random variables with probability density functions (PDFs) $f_X(x)=F_X'(x), f_Y(y)=F_Y'(y)$, then
\be\label{fftprob}
\begin{split}
	\BBP\ls X\ge Y\rs &= \BBE\ls \1(X\ge Y)\rs = \int_\BBR\int_\BBR \! \1(X\ge Y) f_X(x)f_Y(y)\,\dx\dy\\
	& = \int_\BBR \! F_Y(x)f_X(x)\, \dx = \Bigl \langle F_Y,f_X \Bigr \rangle \ .
\end{split}
\ee
In general, the Hermitian inner product on $\BBR$ is defined as $\left\langle f,g \right\rangle=\int_{-\infty}^{\infty} \! \bar f(x) g(x)\,\dx$, but here, both operands are real functions and the conjugate operator disappears.
A second fact is that if $X_1, X_2, \cdots, X_n$ are $n$ independent random variables with PDFs $f_{X_i}$, then the PDF of the sum $X = X_1 + X_2 + \cdots + X_n$ is the convolution
\be
\label{Aequation}
	f_X = f_{X_1} \circledast f_{X_2} \circledast \cdots \circledast f_{X_n} = \circledast_{k=1}^{n} f_{X_k}\ ,
\ee
where the convolution product of two functions is the function defined by $(f\circledast g)(x)=\int_\BBR f(y)g(x-y)dy$. For convolution powers, we write $ \circledast_{k=1}^{n} f_{X}= f_{X}^{\ \circledast n} $\ .

The following cascade mapping provides a closed set of recursive formulas for $n\ge 1$ starting with $p_{jk}^{(0) }=\widetilde p_{jk}^{(0)}, \widehat q_{jk}^{(0) }, t^{(0)}_{kj}=p_{k}^{(0) }$ determined by the initial shock and stress probabilities. 

\noindent{\bf Cascade Mapping:\ } For any $n\ge 1$, suppose $p_{jk}^{(n-1) },\widehat q_{jk}^{(n-1) },\widetilde p_{jk}^{(n-1)}, t^{(n-1)}_{kj}$ are known. The cascade mapping gives the quantities $p_{jk}^{(n) }, \widehat q_{jk}^{(n) }, \widetilde p_{jk}^{(n)}, t_{jk}^{(n) }$ recursively by
\beq
\label{pnm}
p_{jk}^{(n)} &=& \left\langle D_{jk}, \l(g_{j}^{(n-1)}\r)^{\circledast j} \right\rangle\ , \\
\label{hatpnm}\widetilde p_{jk}^{(n)} &=& \left\langle D_{jk}, \l(g_{j}^{(n-1)}\r)^{\circledast j-1} \right\rangle\ , \\
\label{hatqn}\widehat q_{jk}^{(n)} &=&  \left\langle S_{jk},  \l(h_{k}^{(n-1)}\r)^{\circledast k}\right\rangle\ ,  \\ 
\label{tn} t_{kj}^{(n) }&=& t_{kj}^{(n-1) } +(p^{(n)}_{k}-p^{(n-1)}_{k})(1-\widehat q_{j}^{(n-1) })\ ,\eeq
where $D_{jk}$ and $S_{jk}$ are defined in (\ref{DeltaCDF}) and (\ref{SigmaCDF}), respectively, and we use \eqref{pphatq} to compute $p_{k'}^{(n-1) },\widehat q_{j'}^{(n-1) },\widetilde p_{j'}^{(n-1)}$.
Moreover, the stress probabilities $ q_{jk}^{(n) }$ are determined by
\beq  \label{q}1- q_{jk}^{(n) }- p_{jk}^{(n) } &=&\BBP[v\in\hat\CS^c_n\cap\CD^c_n \mid v\in\CN_{jk}]\nonumber\\
&=&\l(1- \widehat q_{jk}^{(n) }\r)\l(1-\left\langle D_{jk},\l(\tilde g_{j}^{(n-1)}\r)^{\circledast j} \right\rangle\r)\ .\eeq
The probability distribution functions in these formulas are also computed recursively:
 \beq
 g_{j}^{(n-1)}(x)&=&\sum_{k'}\Bigl[(1-p^{(n-1)}_{k'})\delta_0(x)+t^{(n-1)}_{k'j}w_{k'j}(x)\nonumber \\&&+(p^{(n-1)}_{k'}-t^{(n-1)}_{k'j})\cdot\frac1{1-\lambda}w_{k'j}(x/(1-\lambda))\Bigr]\cdot  Q_{k' \mid j}
\label{gcdf}\ , \\
h_{k}^{(n-1)}(x)&=&
\sum_{j'} \Big[(1-\widehat q^{(n-1)}_{j'})(1-\widetilde p^{(n-1)}_{j'}))\delta_0(x)+\widetilde p^{(n-1)}_{j'}w_{kj'}(x)\nonumber \\
&&+ \widehat q_{j'}^{(n-1) }(1-\widetilde p^{(n-1)}_{j'})
\cdot\frac1{\lambda}w_{kj'}(x/\lambda)\Big]\cdot Q_{j' \mid k} \ ,\label{hcdf}\\
\tilde g_{j}^{(n-1)}(x)&=&\sum_{k'}\Bigl[(1-p^{(n-1)}_{k'})\delta_0(x)+p^{(n-1)}_{k'}w_{k'j}(x)\Bigr]\cdot  Q_{k' \mid j}
\eeq
with $Q_{k \mid j}=\frac{Q_{kj}}{Q_j^-}, Q_{j \mid k}= \frac{Q_{kj}}{Q_k^+}$.

\noindent {\bf Justification:\ } To justify \eqref{pnm} we need to suppose that the collection of random variables $\Omega_{wv}\xi^{(n-1)}_{wv}$ for different $w\in\CN^-_v$ are mutually conditionally independent. However, there are two sources of dependence: the first is the usual breaking of the LTI property for finite $N$ due to cycles in the skeleton, the second, specific to this model, is the dependence on whether $v$ is in $\CS_{n-1}$. Our approximation is to neglect both sources of dependence, allowing the use of \eqref{fftprob} to give
\beq p_{jk}^{(n)} &=&\BBP[\Delta_v\le  \sum_{w\in \Nvm} \Omega_{wv}\xi^{(n-1)}_{wv} \mid v\in\CN_{jk}]
\nonumber\\
&=&\left\langle D_{jk}, \l(g_{j}^{(n-1)}\r)^{\circledast j} \right\rangle\ .
\eeq
where
\be
g_{j}^{(n-1)}(x)=\sum_{k'}Q_{k' \mid j}\frac{d}{dx}\BBP[\Omega_{wv}\xi^{(n-1)}_{wv}\le x \mid v\in\CN_{jk}, w\in\Nvm, k_w=k'] \ .
\ee 
Under the conditions $v\in\CN_{jk},w\in\Nvm, k_{w}=k'$, the events\\ $\{\xi^{(n-1)}_{wv}=0\},\ \{\xi^{(n-1)}_{wv}=1\}$,$
\{\xi^{(n-1)}_{wv}=1-\lambda\}$ have conditional probabilities\\ $1-p^{(n-1)}_{k'}$, $t^{(n-1)}_{k'j}$, $(p^{(n-1)}_{k'}-t^{(n-1)}_{k'j})$ respectively and the three events are conditionally independent of $\Omega_{wv}$ assuming the LTI property. These facts lead to \eqref{gcdf}. 

To verify \eqref{hatpnm}, we use \eqref{default_WOR} instead of \eqref{Dndef} and follow these same steps. To verify \eqref{hatqn}, we use \eqref{stress}, \eqref{hatqdef} and \eqref{fftprob} to give the formula
\be
\widehat q_{jk}^{(n)} =\BBP[\Sigma_v\le  \sum_{w\in \Nvp} \Omega_{vw}\hat\zeta^{(n)}_{vw} \mid v\in\CN_{jk}]
=\left\langle S_{jk}, \l(h_{k}^{(n-1)}(x)\r)^{\circledast k} \right\rangle
\ee
where
\be
h_{k}^{(n-1)}(x)=
    \sum_{j'}   \frac{d}{dx}\BBP[\Omega_{vw}\hat\zeta^{(n-1)}_{vw}\le x \mid v\in\CN_{jk},w\in\Nvp, j_w=j']\ Q_{j' \mid k}\  .
    \ee
To verify \eqref{hcdf}, note that under the conditions  $k_v=k,w\in\Nvp, j_{w}=j'$, the events $\{\hat\zeta^{(n-1)}_{vw}=0\},\{\hat\zeta^{(n-1)}_{vw}=1\},
\{\hat\zeta^{(n-1)}_{vw}=\lambda\}$  are equivalent to the events\\ $\{w\in\hat\CS^c_{n-1}\cap\CD_{n-1}^c\WOR v\}, \ \{w\in\CD_{n-1}\WOR v\},\{w\in\CS_{n-1}\}$ and hence have conditional probabilities $(1-\widehat q^{(n-1)}_{j'})(1-\widetilde p^{(n-1)}_{j'}),\ \widetilde p^{(n-1)}_{j'}, \widehat q_{j'}^{(n-1) }(1-\widetilde p^{(n-1)}_{j'})$ respectively. \
To verify \eqref{tn}, we apply the LTI property again to compute the conditional probability of the disjoint union $\{\xi^{(n)}_{wv}=1\}=\{\xi^{(n-1)}_{wv}=1\}\cup
\{w\in\CD_n\cap\CD^c_{n-1}, v\in\hat\CS^c_{n-1}\}$ defined by \eqref{xi_def}.

Finally, to verify \eqref{q}, we note that
$\hat\CS^c_n\cap\CD^c_n=\hat\CS^c_n\cap\{ \Delta_v>  \sum_{w\in \Nvm} \Omega_{wv} {\bf 1}_{\{w\in\CD_{n-1}\}}\}.$
By LTI, these last two events are independent conditioned on $v\in\CN_{jk}$, and the required formula results by following the steps taken to prove \eqref{pnm}.

\begin{remark} We see from this argument that the cascade mapping has two sources of error. The first, stemming from cycles in the skeleton that cause finite size corrections to the LTI property, is familiar, and known to behave like $O(N^{-1})$ on configuration graphs as the network size $N$ grows to infinity, and to be zero on finite tree graphs. The second source of error is less familiar and stems from dependence due to the assumption that the default shocks that impact $v$ are diminished by a factor $1-\lambda$ when $v$ becomes stressed, that is, they depend on whether $v$ is stressed or not. 

\end{remark}

The iterates of this cascade mapping converge as $n\to\infty$ to a fixed set of probabilities that represent the eventual state of the system at the end of the cascade. These final probabilities can be used to measure the overall impact of the crisis. For example,
the expected number of eventually defaulted and stressed banks are 
\beq\label{expecteddefaults}
\hspace{-0.7cm}\mbox{Expected number of defaulted banks}&=&N\sum_{jk}P_{jk} \ p^{(\infty)}_{jk}=N\sum_{k}P^+_{k} \ p^{(\infty)}_{k}\ \\
\label{expectedstress}\hspace{-0.7cm}\mbox{Expected number of stressed banks}&=&N\sum_{jk}P_{jk} \ q^{(\infty)}_{jk}\ .
\eeq

\section{Networks with Fixed Skeletons}\label{RWN}

The goal of the present section is to derive approximate probabilistic formulas describing the double cascade on a network where the skeleton is actually known (deterministic) and finite, while the buffers and weights are random. This analysis will allow us to address systemic risk in tractable models of real observed financial networks, without the need for Monte Carlo simulations. 

Let $A=A_{vv'},v,v'\in\CN$ be the nonsymmetric adjacency matrix of the fixed directed graph $(\CN,\CE)$. We number the nodes in $\CN$ by $v=1,2,\dots,N$ and the links by $\ell=1,2,\dots, E$ where $E=\sum_{1\le v,v'\le N} A_{vv'}$. The buffer random variables $\Delta_v,\Sigma_v$ at each node are assumed to have a mass $p^{(0)}_v,q^{(0)}_v$ at $0$ (representing the initial default and stress probabilities) and continuous support  with density functions $d(x),s_v(x)$ on the positive reals. The edge weights $\Omega_\ell, \ell\in\CE$ have continuous support with densities $w_\ell(x)$ on the positive reals but no mass at $0$. The random variables $\{\Delta_v,\Sigma_v,\Omega_\ell\},  v\in\CN, \ell\in\CE$ are assumed to be an independent collection. 

The aim of this section is to use the LTI property as an approximation to derive formulas for the marginal likelihoods $p^{(\infty)}_v,q^{(\infty)}_v$ for the eventual default and stress of all individual nodes, as well as the possibility to compute formulas for more detailed systemic quantities. This approximation is not exact for the same two reasons as before: there is dependence between the default shocks hitting a bank $v$ due to the stress response, and when there are cycles in the skeleton. In general, when the skeleton is a single random realization from a configuration graph ensemble, we expect the LTI approximation to get better with increasing $N$. We now present an approximate analysis, paralleling the previous section, of the sequence of probabilities 
\be
\begin{split}
p^{(n)}_v&=\BBP[v\in\CD_n]\ ,\\
q^{(n)}_v&=\BBP[v\in\CS_n]\ ,\\
\widetilde p^{(n)}_{wv}&=\BBP[v\in\CD_n\WOR w]\ ,\\
\widehat q^{(n)}_v&=\BBP[v\in\hat\CS_n]\ , 
\end{split}\ee
 for each node $v$ or edge $wv$. For the same reason as before we need in addition to track 
\be t^{(n)}_{wv}=\BBP\l[\xi^{(n)}_{wv}=1 \mid  v\in\CN^+_w\r]\ .
\ee

Inductively, we have 
\beq
p_{v}^{(n)} &=& \left\langle D_{v},\circledast_{v' \in \Nvm}  \l(g_{v'v}^{(n-1)}\r) \right\rangle\ , \\
\widetilde p_{wv}^{(n)} &=& \left\langle D_{v}, \circledast_{v' \in \Nvm\setminus w} \l(g_{v'v}^{(n-1)}\r)\right\rangle\ , \\
\widehat	q_{v}^{(n)} &=&  \left\langle S_{v}, \circledast_{v' \in \Nvp}  \l(h_{vv'}^{(n-1)}\r)\right\rangle\ , \\
t^{(n)}_{wv}&=&t^{(n-1)}_{wv}+(p_{w}^{(n)} -p_{w}^{(n-1)} )(1-\widehat	q_{v}^{(n-1)} )\ , \\
q^{(n)}_v&=&1-p_{v}^{(n)}-\left(1-\widehat	q_{v}^{(n)} \right)\left\langle D_{v},\circledast_{v' \in \Nvm}  \l(\tilde g_{v'v}^{(n-1)}\r) \right\rangle\ .\eeq
The PDFs can be computed as before under the LTI approximation:
\beq
\hspace{-4in}g_{wv}^{(n-1)}(x)&=&(1-p^{(n-1)}_{w})\delta_0(x)+t^{(n-1)}_{wv}w_{wv}(x)\nonumber \\&&+\ (p^{(n-1)}_{w}-t^{(n-1)}_{wv})\cdot\frac1{1-\lambda}w_{wv}(x/1-\lambda)\ , \\
h_{vw}^{(n-1)}(x)&=&(1-\widehat q^{(n-1)}_{w})(1-\widetilde p^{(n-1)}_{vw})\delta_0(x)+\widetilde p^{(n-1)}_{vw}w_{vw}(x)\nonumber \\&&+ \widehat q_{w}^{(n-1) }(1-\widetilde p^{(n-1)}_{vw})
\cdot\frac1{\lambda}w_{vw}(x/\lambda)\ ,\\
\tilde g_{wv}^{(n-1)}(x)&=&(1-p^{(n-1)}_{w})\delta_0(x)+p^{(n-1)}_{wv}w_{wv}(x)\ .
\eeq

\section{Numerical Experiments}

In this section, we report briefly on numerical experiments that illustrate the methods developed in this paper.  Firstly, we aim to convince the reader that the LTI method correctly computes the fixed point of the double cascade mapping in networks with large values of $N$ by cross validating it using an independently coded Monte Carlo (MC) implementation. For efficiency, the LTI implementation uses a Fast Fourier Transform approach to compute the convolutions in \eqref{pnm}-\eqref{tn}. This technique was developed in \cite{HurdGlee13} and is sketched in Appendix A.  Secondly, we will show how the LTI method can lead to answers to questions about the nature of systemic risk, particularly the intertwining of stress and default. Thirdly, we shall show how the method performs in a challenging stylized network specified to reflect the complex characteristics of a 2011 dataset on the network of 90 most systemically important banks in the European Union. 

\subsection{Experiment 1: Verifying the LTI Method}

This experiment aims to verify that the LTI method performs as expected when applied to a stylized financial network whose specification is similar to that given in \cite{GaiKapa10a}. It consists of a random directed Poisson skeleton with $N =20000$ nodes and mean degree $z=10$, where each node $v$ can be viewed as a bank with a default buffer $\Delta_v= 0.04$ and stress buffer $\Sigma_v=0.035$. Unlike the deterministic interbank exposures used in \cite{GaiKapa10a}, the weight $\Omega_{\ell}$ of an edge $\ell$ is taken from a log-normal distribution with mean $\mu_\ell=0.2j_\ell^{-1}$, and standard deviation $0.383\mu_\ell$. Note that this specification makes the exposure size dependent on the lending bank. An initial shock is applied to the network that causes each bank to default with a $1\%$ probability.

We compare the final fractions of defaulted and stressed banks as computed using MC simulation with 1000 realizations and the LTI analytic formulas in \eqref{expecteddefaults} and \eqref{expectedstress}. It is simple to generate a directed Poisson random graph of size $N$ with mean degree $z>0$: one simply selects directed edges independently from all $N(N-1)$ potential edges, each with probability $p=z/(N-1)$. The resultant bi-degree distribution is a product of binomials, $P_{jk}=\BBP[v\in\CN_{jk}]={\rm Bin}(N-1,p,j)\times {\rm Bin}(N-1,p,k)$, which for large $N$ nodes is approximately a product of  ${\rm Poisson}(z)$ distributions. 

\begin{figure}[h]
\centering\includegraphics[width=0.7\columnwidth]{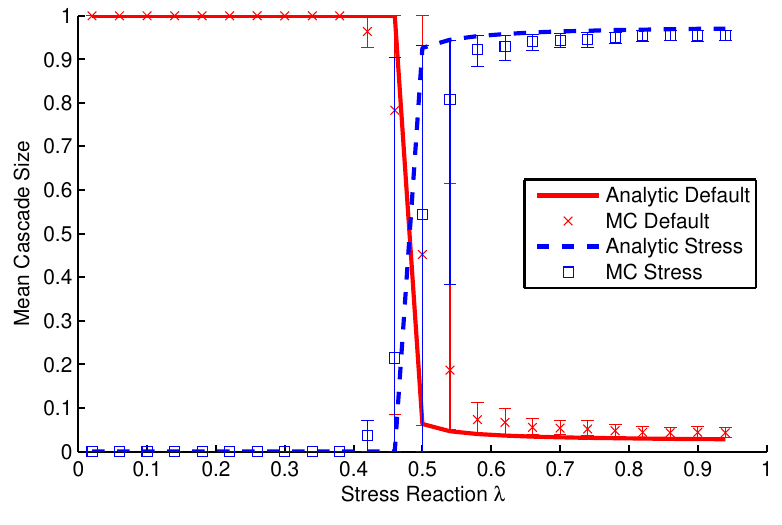}
\caption{Experiment 1. The mean default and stress cascade sizes computed by MC simulations (symbols) and LTI analytics (lines). The skeleton is a Poisson directed network with $N=20000$ nodes and mean degree $z=10$. All banks have default buffers $\Delta_v= 0.04$ and stress buffers $\Sigma_v=0.035$. An edge weight $\Omega_{\ell}$ of an edge $\ell$ is taken from a log-normal distribution with mean $\mu_\ell=0.2j_\ell^{-1}$ and standard deviation $0.383\mu_\ell$. Error bars indicate the $10^{th}$ and $90^{th}$ percentiles of the MC result.}
\label{StylizedGraph}
\end{figure}

Figure \ref{StylizedGraph} plots the results as functions of the stress response parameter $\lambda$, with error bars that represent the $10^{th}$ and $90^{th}$ percentile of the MC result. It shows the expected agreement between MC and LTI analytics, with discrepancies that can be attributed to finite $N$ effects present in the MC simulations.

One fundamental property of our model is clearly shown in this experiment: stress and default are negatively correlated. This fact can be explained by the stress response which enables banks to react to liquidity shocks before they default, by reducing their interbank exposures. This response creates yet more stress, but leads to a more resilient network.  The ``knife-edge'' property of default cascades is also clearly shown: In the model parametrization we chose, a very small increase in $\lambda$ dramatically alters the stability of the network. We also note that MC error bars are very large near the knife-edge.  

\subsection{Experiment 2: A Stylized Poisson Network}
The next experiment focuses again on Poisson networks, with the aim to better understand the effects of various parameters on network resilience. In general, we continue to find confirmation that the LTI results accurately reflect observations from MC simulations.

\subsubsection{Experiment 2A: Effects of Default and Stress Buffers}
We consider how the parametrized financial network of Experiment 1 in a default-susceptible state with $\lambda=0.5$ can be made resilient to random shocks by either increasing the default buffers or decreasing the stress buffers. Such changes, for example, can be prompted by financial regulators.

\begin{figure}[b!]
\centering
\includegraphics[width=0.47\columnwidth]{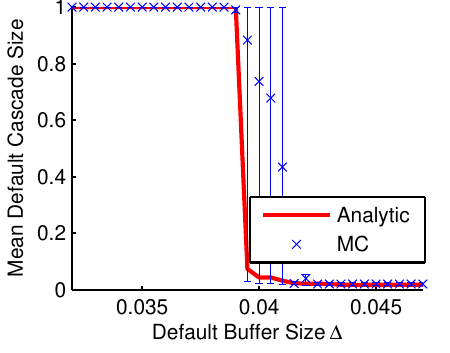}
\put(-180,150){\text{\bf({a})}}
\includegraphics[width=0.47\columnwidth]{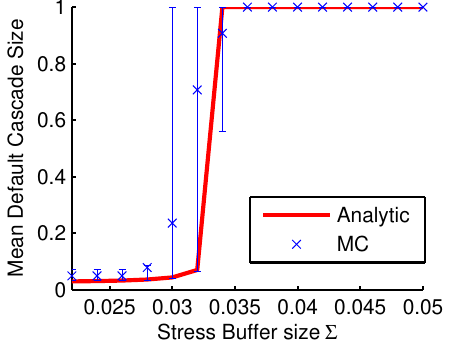}
\put(-180,150){\text{\bf({b})}}
\caption{ Experiment 2A. The mean default cascade size as a function of (a) default buffer $\Delta$ and (b) stress buffer $\Sigma$. The LTI analytic approximation (lines) correctly predicts MC simulations results (symbols). Here $\lambda=0.5$ and other parameters are chosen as in Experiment 1 (Fig.~\ref{StylizedGraph}).}
\label{BuffersGraph}
\end{figure}

In Fig.~\ref{BuffersGraph}(a) we illustrate how the change in default buffers affects the default cascade size. We observe a very fast transition from 100\% default cascade to almost no default as $\Delta$ increases over the interval $[0.04,0.045]$. This knife-edge property is observable in both the LTI analytics and in the MC simulations. Note again that the MC error bars, representing the $10^{th}$ and $90^{th}$ percentiles, become very large near the knife-edge.   

In Fig.~\ref{BuffersGraph}(b), we examine the influence of the stress buffer on the default cascade size. If stress buffers are reduced, banks start to react to stress shocks more quickly, which can in turn dramatically reduce default cascade risk in the network. Figures~\ref{BuffersGraph}(a) and (b) show that there may be several approaches to improving network resilience.

\subsubsection{Experiment 2B:  Effects of Graph Connectivity and Stress Response}
Aside from mandating changes to the behaviour of banks during or prior to a crisis by imposing constraints on stress and default buffers, regulators can also influence the shape of the financial network as a whole. In Experiment 2B, we demonstrate how systemic risk is influenced by the skeleton itself. To this end, we calculate the sizes of default and stress cascades in a directed Poisson network as a function of the connectivity parameter $z$ and the stress response $\lambda$. In our simple model specification, the mean degree $z$ is the only parameter that controls the shape of the skeleton, whereas in a more realistic modelling approach the skeleton may have many more parameters.

In this experiment, we increased the model complexity by assuming each bank has a random default buffer taken from a log-normal distribution with mean $0.18$ and standard deviation $0.18$, and a stress buffer from an independent log-normal distribution with mean $0.12$ and standard deviation $0.12$. The edge weights $\Omega_\ell$ come from a log-normal distribution with mean and standard deviation proportional to $(j_\ell k_\ell)^{-0.5}$, with the average edge weight on the entire network equal to $1$. Once again we apply an initial shock so that each bank has $1\%$ chance of defaulting initially. 

\begin{figure}[!htbp]
\centering
\includegraphics[width=0.47\columnwidth]{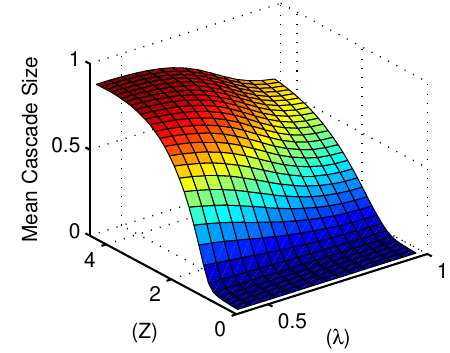}
\put(-180,160){\text{\bf({a})}}
\includegraphics[width=0.47\columnwidth]{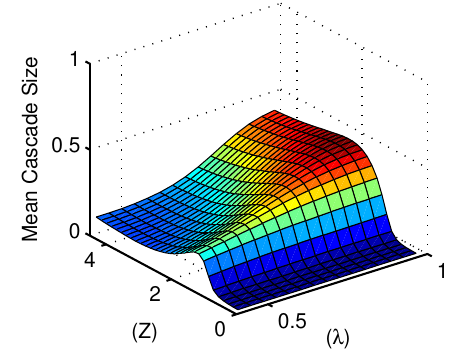}
\put(-180,160){\text{\bf({b})}}
\caption{Experiment 2B. The sizes of the mean default cascade (a) and the mean stress cascade (b) on a directed Poisson network as functions of the network mean degree $z$ and stress response parameter $\lambda$. Here the values of edge weights (interbank exposures), default and stress buffers are taken from log-normal distributions as specified in the text. Other parameters are the same as in Fig.~\ref{StylizedGraph}.}
\label{SurfacePlot}
\end{figure}

In Figs.~\ref{SurfacePlot}(a) and (b), we respectively show the mean sizes of default and stress cascades as functions of network mean degree $z$ and the stress response $\lambda$. For clarity of the graphics, we do not show the MC simulations, as they agree, in a similar fashion as earlier, with the LTI analytics . Again, in these plots we notice the strong anti-correlation between stress and default probabilities as we vary $z$ and $\lambda$. It is also interesting to observe that the final level of stress is not monotonic in the connectivity parameter $z$.

\subsection{Experiment 3: An EU-Inspired Network with 90 Nodes}\label{sec6.3}

It is well known that Poisson random graphs are an inadequate description of real economic networks, so we are interested in having a rough picture of the systemic risk of actual financial networks. In Experiment 3, we consider a single realization of a 90 node graph that aims to capture stylized features of the interbank network of the European Union. We computed cascade dynamics on this network using both the MC simulations and the LTI analytic method of Section \ref{RWN} for networks with a fixed skeleton. As a preliminary step (not reported), we validated the consistency of the LTI analysis by verifying that the two methods agree as expected on a number of tree networks.

\begin{figure}[!htbp]
\centering\includegraphics[width=0.47\columnwidth]{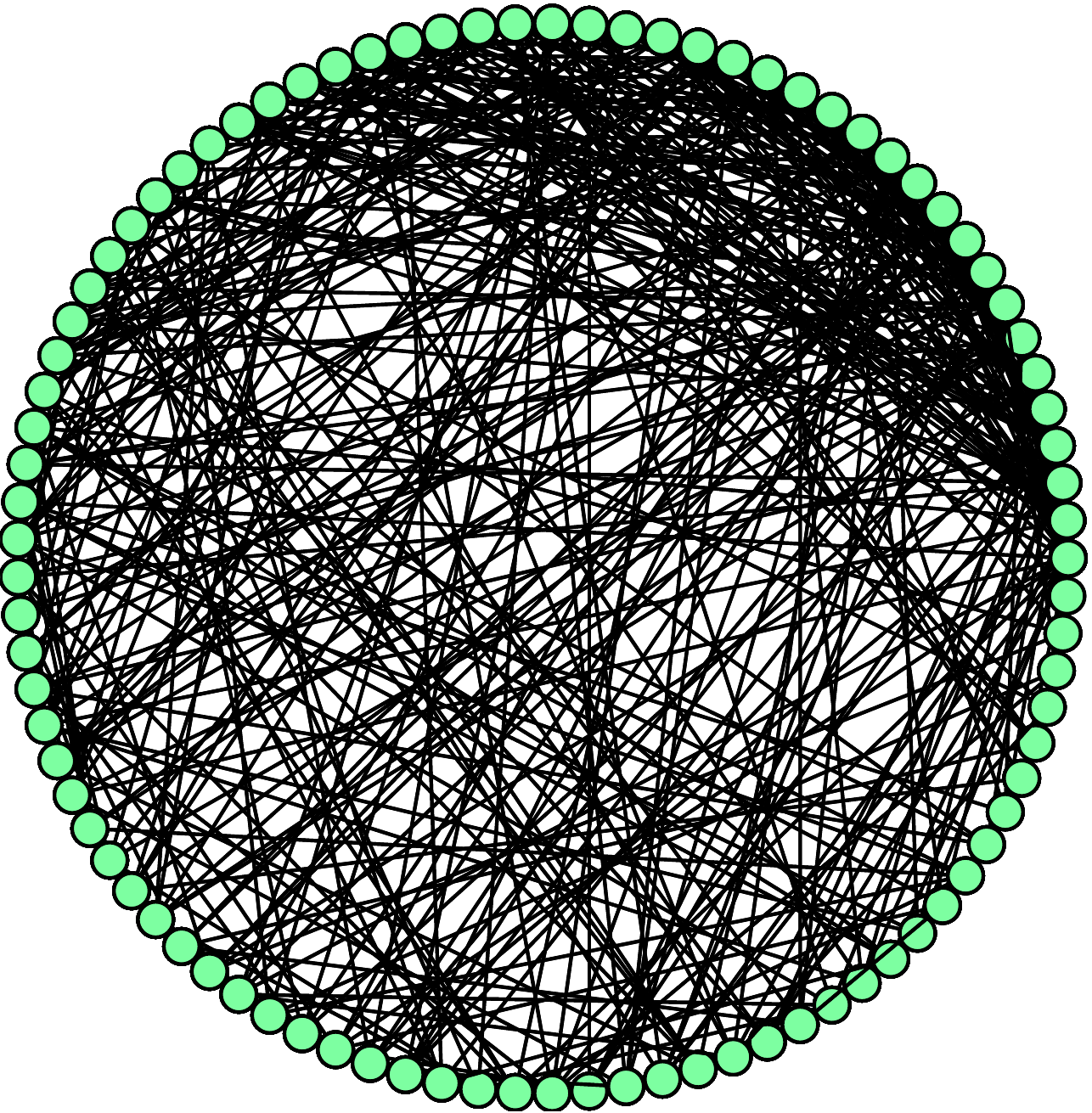}
  \caption{A representation of the skeleton of the 90 bank network of Experiment 3. We do not show the edge directions here to avoid cluttering the figure. The nodes are plotted with total degree increasing in the clockwise direction, with the minimally connected bank being the rightmost node.}
  \label{EUnetwork}
\end{figure}

Numerous studies of real-world financial networks, notably \cite{BechAtal10} and \cite{ContMousSant13}, have observed their  highly heterogeneous structure and concluded that in and out degrees have fat tailed distributions, as do exposure sizes, and presumably, buffers. Our schematic model of 90 EU banks was designed to capture these basic statistical features, and as well to fit aggregated statistics from data published on the 2011 ECB stress testing of systemically important banks in the European Union. We show the skeleton of our stylized network in Figure~\ref{EUnetwork}. The details of its construction and the specifications for buffer and exposure distributions are given in Appendix B.

\begin{figure}[!htbp]
\centering
\includegraphics[width=0.47\columnwidth]{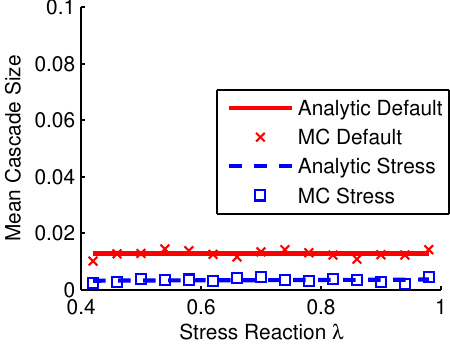}
\put(-30,140){\text{\bf({a})}}
\includegraphics[width=0.47\columnwidth]{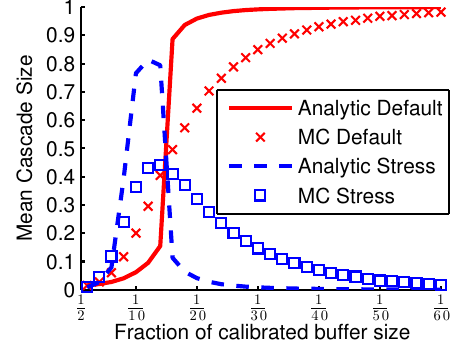}
\put(-30,140){\text{\bf({b})}}
\caption{Experiment 3: sizes of default and stress cascades on a stylized EU interbank network starting from the default of a single randomly chosen bank. (a) Cascade sizes for various values of the stress response $\lambda$. The EU financial system at the time of the 2011 stress testing exercise appears to be resilient to single-bank shocks. (b) The same system as in (a) where a dire pre-shock crisis has reduced the bank default buffers to 10\% of their original value, and stress buffers to a fraction of their value indicated on the $x$-axis.}
\label{FiniteGraph}
\end{figure}

In the first part of the experiment, we calculate the mean sizes of stress and default cascades that start from the default of one random bank in the EU network. Figure \ref{FiniteGraph}(a) shows the mean default and stress cascade sizes as functions of the stress response parameter $\lambda$. This graph shows no evidence of cascading (beyond the initially defaulted bank and the stress it causes to its immediate neighbors), demonstrating that the MC and analytic computations agree that the EU network in June 2011 was resilient to such a shock.

To move the EU network to a knife-edge situation where a large scale double cascade can be triggered by the default of a single bank, we found it was necessary to imagine a dire crisis where prior to the default shock, the Core Tier 1 Capital of all institutions (i.e. their default buffers) has been decimated to 10\% of their initial amount. Furthermore, each bank's stress buffer is also reduced to a certain fraction of its original value. In Fig.~\ref{FiniteGraph}(b) we take $\lambda = 0.7$ and show the cascade sizes versus the remaining fraction of stress buffers.

The MC simulations show that for smaller shocks the size of a stress cascade first dominate default cascades: the banking system becomes highly illiquid, but most banks are able to protect themselves from default. Beyond a certain critical value of the global shock, however, default cascades take over stress cascades, as banks are hit by a shock so high that they default immediately, without the possibility to protect themselves as they would if they got first in the stressed state. Our analytical approach is able to capture the essence of these mechanisms, identifying correctly the critical point where default cascades overtake stress cascades. The quantitative size of the cascades away from this critical point, however, is less well matched by the analytical model, probably because of the presence of short cycles in the skeleton. As cycles tend to slow down the cascades, the transitions are smoother in the MC. It is quite interesting that, in spite of our two approximations, our model is able to capture the general mechanisms: a peak in the number of stressed banks is a signal that can indicate quite precisely the point where systemic default takes over in a banking network.
 
\section{Conclusions}

Our double cascade model is a natural extension of a strand of systemic risk research that studies elementary models that build in either default or stress cascades, but not both. Only by combining the default and stress mechanisms into a single model can one measure such features as the intuitively obvious effect of banks using the stress response to reduce their risk of default. 

Developing a feasible and reliable computation framework for a model as complex as ours poses challenges that have been met in the experiments described in Section 5. We have demonstrated how computations can be done by two complementary approaches: the Monte Carlo (MC) method and the locally tree-like independence (LTI) analytic method. Having these two approaches, each with its pros and cons, allows cross validation,  increasing confidence in the results one obtains. MC simulation, with its natural flexibility, remains the workhorse computational framework for general systemic network risk.   However, where it applies, the LTI method, given its relation to similar methods in other areas of network science, adds the possibility to understand the flow of the cascade in a different and sometimes better way than one can with MC simulations alone. For example, using LTI one can determine sensitivities to changing parameters through explicit differentiation. In some situations, for example in simulating general assortative $(P,Q)$ configuration models, the MC method is infeasible or leads to unacceptably long run times,  whereas the LTI method can be computed without difficulty. Ultimately what is important is that both methods have complementary strengths and weaknesses, and when used in combination lead to robust and reliable conclusions about a wide range of network effects.

Our numerical experiments explore only a small sample of simple model specifications, leaving many promising financial networks to be investigated using our techniques. While the systemic importance of parameters such as network connectivity, mean buffer strength, and the size of the interbank sector have been studied previously, other  parameters such as the stress response, the  buffer and exposure variances, and graph assortativity, remain almost completely unexplored. The effect of market illiquidity and asset fire sales has been omitted from the present paper, but its impact on the cascade and consequently the greater economy merits careful investigation. Financial network databases, and the statistical methods for matching such data to the model, are still in an underdeveloped state, but are needed to tie down the wide range of parameters in our model. Planned future investigations of the double cascade model and its extensions will uncover and explain further interesting and unexpected systemic risk phenomena, and find uses by policy makers and regulators.

\begin{appendices}

\section{Discrete Probability Distributions and the Fast Fourier Transform}

Numerical implementation of these models follows the methods outlined in \cite{HurdGlee13}. In this section, we analyze the case where all random variables $\{\Delta_v,\Sigma_v,\Omega_\ell\}$ take values in a specific finite discrete set $\CM=\{0,1,\dots,(M-1)\}$ with a large value $M$. It is well known that in this situation the convolutions in \eqref{Aequation}  are slow to compute by direct integration, but can be performed exactly and efficiently by use of the discrete Fast Fourier Transform (FFT). 

Let $X,Y$ be two independent random variables with probability mass functions (PMF) $p_X,p_Y$ taking values on the non-negative integers $\{0,1,2,\dots\}$. Then the random variable $X+Y$ also takes values on this set and has the probability mass function (PMF)
$ p_{X+Y}=p_X*p_Y$ where the convolution of two functions $f,g$ is defined to be
\be\label{convolution} (f*g)(n) = \sum_{m=0}^n f(m) g(n-m)\ .\ee
Note that $p_{X+Y}$ will not necessarily have support on the finite set $\CM$ if  $p_X,p_Y$ have support on $\CM$. This discrepancy leads to the difficulty called ``aliasing''. 

We now consider the space  $\BBC^M$ of $\BBC$-valued functions on $\CM=\{0,1,\dots,M-1\}$. The discrete Fourier transform, or fast Fourier transform (FFT), is the linear mapping $\CF:a=[a_0,\dots,a_{M-1}]\in\BBC^M\to \hat a=\CF(a)\in \BBC^M$ defined by
\be \hat a_k=\sum_{l\in\CM} \zeta_{kl} a_l\ , k\in\CM\ . \ee
where the coefficient matrix $Z=(\zeta_{kl})$ has entries $\zeta_{kl}=e^{-2\pi i kl/M}$. 
The ``inverse FFT'' (IFFT), is given by the map $a\to\tilde a=\CG(a)$ where 
\be\tilde a_k=\frac1{M}\sum_{l\in\CM} \bar\zeta_{kl} a_l\ , k\in\CM\ . \ee
If we let $\bar a$ denote the complex conjugate of $a$, we can define the Hermitian inner product between 
\be  \langle a,b\rangle:=\sum_{m\in\CM} \bar a_m b_m\ .\ee
We also define the convolution product of two vectors:
\be (a*b)(n)= \sum_{m\in\CM}a(m)\ b(n-m\mbox{ modulo $M$} ), \quad n\in\CM \ .\ee
Note that this agrees with \eqref{convolution} if and only if the sum of the supports of $a$ and $b$ is in $\CM$. Otherwise the difference is called an aliasing error: our numerical implementations reduce or eliminate aliasing errors by taking $M$ sufficiently large. 

The following identities hold for all $a,b\in \BBC^M$: \begin{enumerate}
  \item Inverse mappings:  $a=\CG(\CF(a))=\CF(\CG(a))$\ ; 
  \item Conjugation:   $\overline{\CG(a)}=\frac1{M}\CF(\bar a)$\ ; 
  \item Parseval Identity: $ \langle a,b\rangle=M\langle \tilde a,\tilde b\rangle=\frac1{M}\langle \hat a,\hat b\rangle$\ ; 
  \item  Convolution Identities:  $\tilde a\dotstar\tilde b=\widetilde{(a*b)},\ \hat a\dotstar\hat b=\widehat{(a*b)}$\ , 
 \end{enumerate}
where $\cdot*$ denotes the component-wise product. 

As an example to illustrate how the above formulas help, we observe that a typical formula \eqref{pnm} can be computed instead by the formula
\be p^{(n)}_{jk}=\frac1{M}\ \langle \CF(D), \left(\hat g_j^{(n-1)}\right)^j\rangle \ , \ee
where $\hat D=\CF(D),\ \hat g_j^{(n-1)}=\CF(g_j^{(n-1)})$ and the power is the component-wise vector multiplication. Such FFT-based formulas can be computed systematically, very efficiently, if the discrete probability distributions for $\Delta,\Sigma,\Omega$ are initialized in terms of their Fourier transforms.

\section{EU Network Construction}
 
In 2011, the European Banking Authority (EBA) made public a dataset\footnote{http://www.eba.europa.eu/risk-analysis-and-data/eu-wide-stress-testing/2011}  on the  interbank exposures of a selection of 90 medium to large European banks, as well as other information such as their  Core Tier 1 Capital and Risk Weighted Assets.  In this dataset, interbank exposures are aggregated by country on the liability side, which means we only know the aggregated amount each bank $v$ has lent to all banks in each EU country $c$. In this Appendix we explain how we built the synthetic network in Experiment 3 that mimics stylized facts of  real financial networks and uses this EBA data as a source of additional information about interbank liabilities and bank balance sheets. 

Motivated by the ubiquitous relevance of networks with fat-tailed degree distributions, as reported in papers such as \cite{ContMousSant13,BechAtal10}, we built the skeleton based on the preferential attachment model of \cite{BolBorChaRio03}. Given four parameters $\alpha$, $\gamma$, $\delta_{-}$ and $\delta_{+}$ and  letting $\beta=1-\alpha-\gamma$, this model grows a random directed network from a finite initial ``seed graph'' using three rules:
\begin{enumerate}
	\item With probability $\alpha$, add a new vertex $v$ together with an edge from $v$ to an existing vertex $w$, where $w$ is chosen according to $j_w + \delta_{-}$ (that is, with probability proportional to $j_w + \delta_{-}$).
	\item With probability $\beta$, add an edge from an existing vertex $v$,  chosen according to $k_v + \delta_{+}$, to an existing vertex $w$, chosen according to $j_w + \delta_{-}$.
	\item With probability $\gamma$, add a new vertex $v$ together with an edge from an existing vertex $w$ to $v$, where $w$ is chosen according to $k_w + \delta_{+}$.
\end{enumerate}
This method is known to lead to fat-tailed degree distributions $P^-_j \sim j^{-\tau_-}$ and $P^+_k \sim k^{-\tau_+}$ with Pareto exponents $\tau_-=1+\frac{1+\delta_{-}(\alpha+\gamma)}{\alpha + \beta},
\	\tau_{+}=1+ \frac{1+\delta_{+}(\alpha+\gamma)}{\gamma + \beta}.$
The four parameters  $ \alpha=0.169$, $\gamma=0.169$, $\delta_{-}=\delta_{+}=4.417$ are determined by the following conditions: we assumed Pareto exponents $\tau_{-}=\tau_{+}=4$ (ensuring finiteness of certain moments),  that $\alpha=\gamma$, and that the mean degree is $z=10$ (which is a typical value observed in financial networks). To achieve this, we searched over a range of values of $\alpha$,  generating for each $\alpha$ 1000 samples of networks each consisting of $N=1000$ nodes. For each realized network sample, we selected the subnetwork of the 90 most connected nodes and calculated its mean degree $z$. Finally we selected the $\alpha$ value which provided a mean degree $z$ closest to $10$. This set of parameters was then used to produce the skeleton of Experiment 3 by generating a single sample of a (directed) scale-free graph with $N = 1000$ nodes, and retaining its most connected subnetwork of 90 nodes.

In order to build bank balance sheets, we assumed an LTI compatible specification of the buffer and exposure random variables as log-normally distributed conditionally on the network topology:
\beq
	\Delta_v &=&(k_v j_v)^{\beta_1} \exp[a_1+b_1 X_v]\ ,
	\label{delta-calibration}\\
\Sigma_v &=&\frac23 (k_v j_v)^{\beta_1} \exp[a_1+b_1 \tilde X_v]\ , \\
	\Omega_v &=& (k_v j_v)^{\beta_2} \exp[a_2+b_2 X_\ell].
	\label{omega-calibration}
\eeq
where the collection $\{X_v,\tilde X_v,X_\ell\}$ consists of independent standard normal random variables. To fix the parameter values, first we arbitrarily set $ \beta_1=0.3$ and $\beta_2=-0.2$, with the rationale  that the default buffer should increase with  bank connectivity, while a larger number of counterparties should imply lower average bilateral exposures. We used the reported Core Tier 1 Capital as a proxy for the default buffers, and thus we matched the first and second sample moments $E[\Delta_v]$ and $E[\Delta^2_v]$ using equation \eqref{delta-calibration}. Since we found no proxy in the data for the stress buffers $\Sigma$, we arbitrarily selected the same parameters as for $\Delta$, but with a prefactor $2/3$. Finally, matching equation \eqref{omega-calibration} with sample moments  $E[A^{IB}_v]$ and $E[(A^{IB}_v)^2]$, from the aggregated interbank exposure data, gives us enough equations to determine the full list of parameters: $ \beta_1=0.3$, $a_1=8.03$, $b_1=0.9$, $\beta_2=-0.2$, $a_2=8.75$, $b_2=1.16 $.
\section{Acknowledgements} This project was germinated at the MITACS Canada International Focus Period devoted to Advances in Network Analysis and its Applications, organized by Evangelos Kranakis and held in Vancouver, Canada in July 2012. It was funded by awards from the Global Risk Institute for Financial Services in Toronto (TRH), from the Natural Sciences and Engineering Research Council of Canada (TRH), from the Irish Research Council co-funded by Marie Curie Actions under FP7 (INSPIRE fellowship, SM), from the FET-Proactive project PLEXMATH (DC), and from the Science Foundation Ireland (11/PI/1026, DC and SM)

We are  indebted to Lionel Cassier of \'Ecole Polytechnique France, and Huibin Cheng, Matheus Grasselli and Bernardo Costa Lima of McMaster University, Canada, who were active in developing an early version of our model. We are grateful to Grzegorz Halaj of the European Central Bank, for  discussions about the EU financial network.
\end{appendices}
\bibliographystyle{abbrvnat}



\end{document}